\documentclass{aa}  

\usepackage{xcolor}
\usepackage{graphicx}
\usepackage{txfonts}
\usepackage{hyperref}
\newcommand{\hi}{H\,{\sc i}}
\newcommand{\hii}{H\,{\sc ii}}

\begin{document}

   \title{The complex history of NGC\,1427A revealed by its star clusters and star formation history}
\titlerunning{The complex history of NGC\,1427A revealed by its star clusters and SFH}

   \subtitle{}

   \author{Katja Fahrion
          \inst{1}
          \and
          Michael Hilker
          \inst{2}
          \and
          Avinash Chaturvedi
          \inst{3}
          \and
          Juan P. Carvajal
          \inst{4}
          \and
          Thomas H. Puzia
          \inst{4}
          }

   \institute{Department of Astrophysics, University of Vienna, T\"{u}rkenschanzstra{\ss}e 17, 1180 Wien, Austria\\
             \email{katja.fahrion@univie.ac.at}
             \and
             European Southern Observatory, Karl-Schwarzschild-Stra{\ss}e 2, 85748 Garching bei M\"unchen, Germany
             \and
             Leibniz-Institut für Astrophysik Potsdam (AIP), An der Sternwarte 16, 14482 Potsdam, Germany
             \and
             Institute of Astrophysics, Pontificia Universidad Católica de Chile, Av. Vicuña Mackenna 4860, 7820436 Macul, Santiago, Chile
             }

   \date{\today}

 
  \abstract
   {Star-forming low-mass galaxies in the dense environments of galaxy clusters provide opportunities to study how environmental effects such as ram-pressure stripping, tidal interactions, or galaxy mergers shape a galaxy's star formation history. We combined integral-field spectroscopic observations with the Multi Unit Spectroscopic Explorer (MUSE) and available multi-band imaging of the star-forming galaxy NGC\,1427A, located near the centre of the Fornax galaxy cluster, at a distance of 20 Mpc. 
   Our aim was to trace the evolutionary history of NGC\,1427A using the star formation history reconstructed from the integrated spectra and employing star clusters as surviving tracers of past star formation episodes. We fitted the spectral energy distribution of 222 star cluster candidates using archival $u,g,r$, and $i$ photometry to derive the ages and masses. For 58 clusters, we additionally incorporated their MUSE spectra in the fits and found an encouraging agreement between the photometric and spectroscopic results. 
   The comparison of the age distribution of star clusters with star formation histories from a full spectrum fitting of the MUSE data found a reasonable agreement, with evidence for multiple episodes of star formation throughout the history of NGC\,1427A. In particular, we found a population of young clusters ($\sim$ 10 Myr) that is located along the star formation edge and within the northern object, and a population of intermediate-age clusters ($\sim$ 100 - 300 Myr) with corresponding peaks in the star formation history of NGC\,1427A. We interpret these populations in the context of the orbital evolution of NGC\,1427A in the Fornax cluster and conclude that this galaxy has experienced not only ram-pressure stripping, but also tidal interactions or even a minor galaxy merger. The northern object is likely a regular component of the galaxy.}


   \maketitle
%

\section{Introduction}

Dense environments such as galaxy clusters and massive groups are dominated by early-type galaxies that no longer show active star formation in their stellar populations, whereas star-forming late-type galaxies are mostly found in low-density environments such as the field, low-mass groups, or the outskirts of galaxy clusters. This so-called morphology density relation has been established for many decades (e.g. \citealt{Oemler1974, Dressler1980, Whitmore1993, Fasano2000, Fasano2015, Shimakawa2021, PerezMillan2023}) and implies that the environment greatly affects the evolution of galaxies, in particular in terms of when and how star formation proceeds. 

Several phenomena have been proposed that can alter the morphologies, gas content, and consequently, the star formation history (SFH) of galaxies in dense environments, such as ram-pressure stripping (e.g. \citealt{GunnGott1972}), galaxy mergers (e.g. \citealt{Zwicky1951}), or tidal interactions \citep{GallagherOstriker1972, Moore1996}. Ram-pressure stripping refers to the stripping of gas from a galaxy by hydrodynamic interactions with the intracluster medium (see \citealt{Cortese2021} and \citealt{Boselli2022} for recent reviews). This process has been studied in detail with observations (e.g. \citealt{Poggianti_GASP_2017}) as well as simulations (e.g. \citealt{Yun2019}). It was shown to be able in particular to remove neutral gas from a galaxy, while the molecular gas remains and is consumed by ram-pressure-induced star formation, which ultimately shuts star formation down. Similarly, tidal interactions or even mergers with galaxies can trigger strong bursts of star formation while simultaneously removing hot gas (e.g. \citealt{Balogh2000}). This quickly depletes the galaxy gas reservoir. 

The shallow gravitational potentials of low-mass galaxies make them particularly susceptible to such processes, and hence, dwarf galaxies offer a unique laboratory to explore these phenomena and their effect on the SFH of a galaxy (e.g. \citealt{Kenney2014, Stierwalt2015, Boselli2021, Kleiner2023}). We study an especially interesting target that can provide insights into how dense environments shape the star formation and the evolution of low-mass galaxies. NGC\,1427A is an irregular dwarf galaxy in the nearby Fornax galaxy cluster ($D \sim 20$ Mpc; \citealt{Blakeslee2009}). The morphology of this galaxy is peculiar. Its arrow-head shape points towards the south-west, and it has blue colours in optical images (e.g. \citealt{CelloneForte1997}). NGC\,1427A has been at the centre of a long-standing debate about its nature, and ram-pressure stripping, tidal interactions, and galaxy mergers have all been proposed to be the cause of its morphology and properties \citep{CelloneForte1997, Hilker1997, Chaname2000, Georgiev2006, Sivanandam2014, Mora2015, LeeWaddell2018, Mastropietro2021, Serra2024}.

\begin{figure*}
    \centering
    \includegraphics[width=0.85\textwidth]{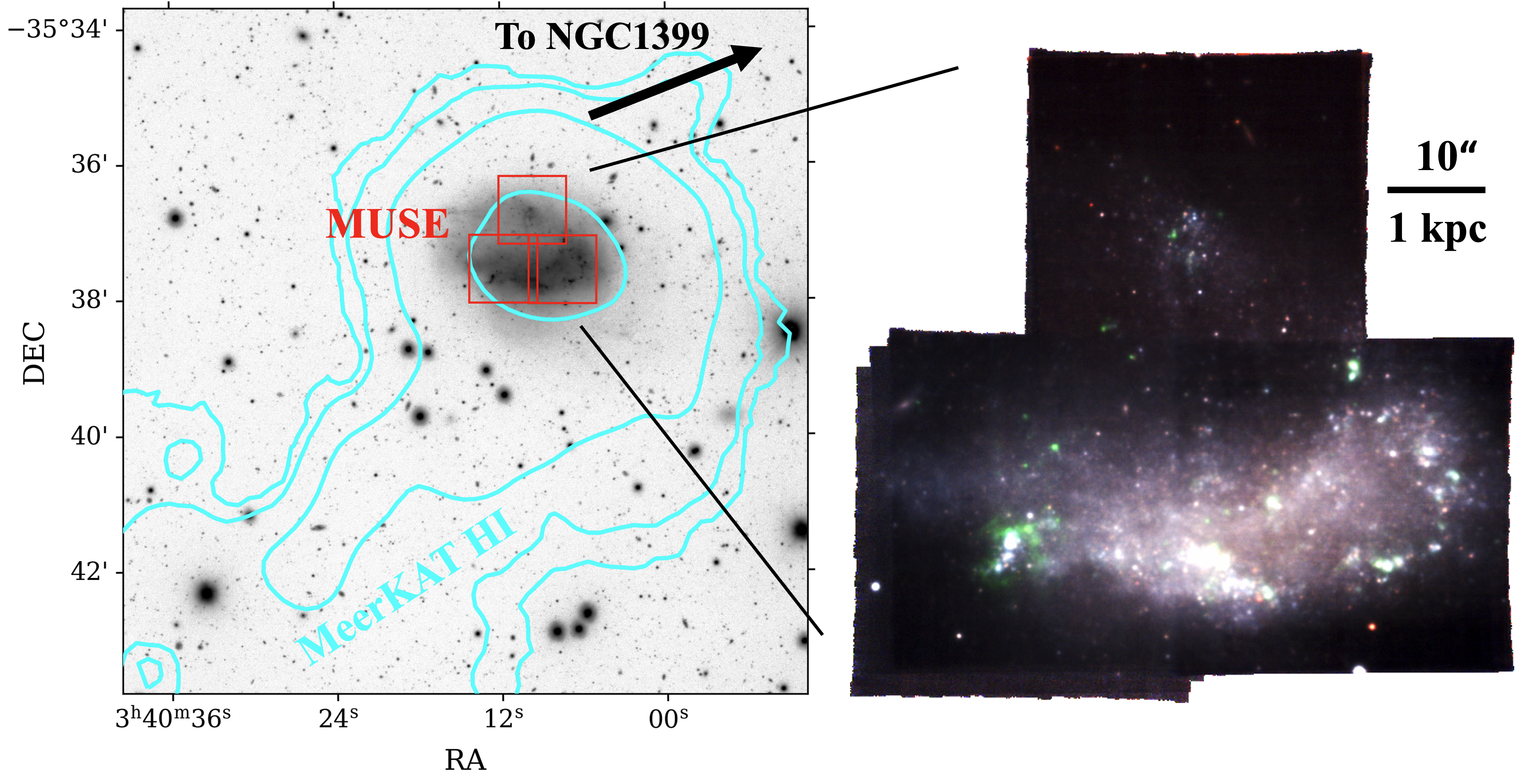}
    \caption{Left: $g$-band image from FDS \citep{FDS} shown in greyscale, illustrating the arrow-head shape of NGC\,1427A, which points towards the south-west. In comparison, the cyan contours show part of the MeerKAT \hi\ contours (from \citealt{Serra2024}, 21\arcsec\,beam size, contours at 1, 10, 100, and 1000 Jy beam$^{-1}$ m s$^{-1}$), which demonstrate a \hi\ tail pointing towards the south-east. The MUSE pointings are indicated with red squares. Right: False-colour image of NGC\,1427A, reconstructed from MUSE data (R: I-band, G: R-band, and B: V-band images created from the MUSE cube). The scale bar refers to 10\arcsec, corresponding to 1 kpc at a distance of 20 Mpc. The MUSE image spans $\sim 2\arcmin \times 2\arcmin$ (12 $\times$ 12 kpc). North is up, and east is to the left.}
    \label{fig:MUSE_colour_img}
\end{figure*}

Based on multi-band optical observations, \cite{CelloneForte1997} described the peculiar morphology of NGC\,1427A as having a distorted, ring-like structure on a diffuse stellar background. They further noted the presence of a clump to the north of the disc, which since then has been denoted as the northern object or northern clump (see Fig. \ref{fig:MUSE_colour_img}). \cite{CelloneForte1997} interpreted the morphology, colours, and the presence of the northern object as a sign of an ongoing merger that possibly triggered the evident star formation across NGC\,1427A. This proposition was challenged by \cite{Chaname2000}, who presented long-slit observations of bright \hii\ regions and found a consistent velocity pattern between the main body of NGC 1427A and the northern object. Similarly, \cite{Hilker1997} found the northern object to be embedded in the diffuse stellar light within a symmetrical distribution and proposed it to be an original part of NGC\,1427A. Further, \cite{Hilker1997} identified a wealth of young stellar clusters and \hii\ regions in the southern part of the galaxy and suggested that star formation was triggered within the last few hundred million years, possibly as a consequence of ram-pressure stripping. A similar conclusion was drawn by \cite{Mora2015}, who used deep imaging to study the star cluster population of NGC\,1427A and proposed that the most recent star formation episodes occurred $\sim 4$ Myr ago.

\cite{LeeWaddell2018} questioned whether ram-pressure stripping might have caused the morphology of NGC\,1427A. Based on \hi\ observations from the Australia Telescope Compact Array (ATCA), they identified a \hi\ tail pointing towards the south-east, perpendicular to the direction suggested by the young stellar populations in the arrow-head appearance of NGC\,1427A, but away from the centre of Fornax. Instead of ram-pressure stripping, \cite{LeeWaddell2018} proposed that the  \hi\ tail was caused by tidal interactions that also triggered star formation within the stellar body. Additionally, they again suggested that the northern object might be still merging with the main galaxy while already having joined the NGC\,1427A rotating disc. \cite{Mastropietro2021} presented tailored hydrodynamical simulations of NGC\,1427A-like dwarf galaxies (albeit less massive, as pointed out by \citealt{Serra2024}). They were able to reproduce the  \hi\ tail and stellar morphology from a combination of ram-pressure stripping and tidal interactions with the potential of the Fornax cluster. In their scenario, the northern object was simply a star-forming clump.

Recently, \cite{Serra2024} presented deep MeerKAT \hi\ observations of NGC\,1427A. In their high-resolution data, they were able to detect the \hi\ tail out to $\sim 70$ kpc. Based on the complex gas kinematics, length, and width of the \hi\ tail, \cite{Serra2024} proposed that the tail might have started out from a high-speed tidal interaction or galaxy merger that started at least 300 Myr ago, but was then shaped by ram-pressure stripping. 

We built on these previous studies by exploring the evolutionary history of NGC\,1427A through the lens of its stellar and star cluster populations. We employed integral-field spectroscopy in conjunction with optical imaging to study the SFH of NGC\,1427A from its stellar light and its star cluster population. A complementary paper \citep{Carvajal2026} presents a comprehensive analysis of NGC\,1427A by combining integral-field spectroscopic data with data from other wavelengths. We describe the available data in Sect. \ref{sect:data} and our spectroscopic analysis in Sect. \ref{sect:MUSE_analysis}. The analysis of the star cluster properties from fitting of their spectral energy distributions (SEDs) is described in Sect. \ref{sect:SED_fitting} and compared to the stellar population analysis in Sect. \ref{sect:SFH}. We discuss our results in Sect. \ref{sect:discussion} and conclude in Sect. \ref{sect:conclusion}.

\section{Data}
\label{sect:data}
To study NGC\,1427A and its star cluster content, we used data from the Multi Unit Spectroscopic Explorer (MUSE). To complement this spectroscopic dataset, we used publicly available deep photometric catalogues of sources in the Fornax cluster obtained with OmegaCam as part of the Fornax Deep Survey (FDS; \citealt{FDS}).

\subsection{MUSE data}
\label{sect:MUSE_data}
MUSE is an integral-field spectrograph mounted at the Very Large Telescope of the European Southern Observatory on Cerro Paranal in Chile \citep{Bacon2010}. In the wide-field mode, MUSE provides a field of view (FOV) of 1\arcmin $\times$ 1\arcmin\,on the sky with a spatial sampling of 0.2\arcsec\,per pixel. In the wavelength direction, MUSE provides a spectrum in the optical wavelength regime from 4700 to 9300 \AA, sampled at 1.25 \AA. 

We made use of archival MUSE data of NGC\,1427A, obtained in 2015 (programme ID 094.B-0612, PI: Mora), combined with newer data obtained in 2022 (programme ID 110.23VM, PI: Fahrion). The older data cover the western part of the main body of NGC\,1427A, while the newer observations cover the eastern part as well as the northern object. The newer data were taken using the adaptive-optics (AO) system, and the older data were acquired without AO. 

Using MUSE pipeline version 2.8.7 \citep{Weilbacher2020} implemented within \textsc{esorex}\footnote{\url{https://www.eso.org/sci/software/cpl/esorex.html}}, we reduced the individual exposures following the standard prescriptions according to the instrument manual. Because the data from 2015 had no dedicated sky observations, the sky reduction was done from the science exposures themselves. As NGC\,1427A covers large portions of the FOV, the standard parameters erroneously include the extended nebular emission, leading to sky residuals at the position of emission lines. To account for this, we set the \textsc{skymodel\_fraction} and \textsc{skymodel\_ignore} parameters to 0.03 and 0.005 in the \textsc{esorex scipost} recipe, respectively. For the data from 2022, we used the dedicated sky exposures for sky reduction following the standard recipes.  

In the last step, we combined all the individual exposures into a final data cube. 
During the observations of the main body in 2022, several observing blocks had to be restarted and repeated due to changing weather conditions during the observations. As some of the interrupted OBs nonetheless provided useful data, the final data cube has a slightly varying depth across the FOV. Figure \ref{fig:MUSE_colour_img} shows a red-green-blue image of NGC\,1427A created from synthetic V, R, I images obtained from the MUSE data cube. Regions with strong H$\alpha$ emission show green colours.

\subsection{Fornax Deep Survey data from OmegaCam}
\label{sect:OmegaCam_data}
The Fornax cluster was imaged with OmegaCam at the Visible and Infrared Survey Telescope for Astronomy (VISTA) as part of the FDS \citep{FDS, Venhola2018}. The data are publicly available and contain deep images in \textit{u, g, r}, and \textit{i} filters of the central 28 square degrees of Fornax. \cite{Cantiello2020} presented an extensive catalogue of compact sources sources extracted from the OmegaCam data, containing in total 1.7 million sources with \textit{ugri} photometry in the AB magnitude system with aperture corrections already applied. \cite{Cantiello2020} presented an example study focusing on globular clusters, but the full sample also includes other compact sources such as younger star clusters. As it is a photometric sample, contamination from background galaxies and foreground stars is also possible.

We selected all sources in the \textit{ugri} catalogue in a 140\arcsec $\times$ 140\arcsec\,square centred on NGC\,1427A, which covers the MUSE FOV. In this region, the FDS catalogue contains 267 sources. We used these sources as our initial star cluster candidate sample. We note that NGC\,1427A was also observed with the \textit{Hubble} Space Telescope in multiple filters (see e.g. \citealt{Georgiev2006}), but we chose to use the ground-based data instead because of the close match in spatial resolution to the MUSE data and homogeneous photometric sampling across the MUSE mosaic FOV (Fig.~\ref{fig:MUSE_colour_img}). However, we note that the limited spatial resolution can lead to the blending of associations of stars and star clusters.

\section{Spectroscopic analysis}
\label{sect:MUSE_analysis}
In the following, we describe the MUSE data analysis. For this, we used well established methods to constrain the stellar and gaseous content of NGC\,1427A (see e.g. \citealt{Emsellem2022, Bulichi2023}).

\subsection{Binned maps}
\label{sect:binned_maps}
To obtain binned maps from the MUSE data, we use the Data Analysis Pipeline (DAP)\footnote{\url{https://gitlab.com/francbelf/ifu-pipeline}}, which aims to obtain high-level science products from integral-field spectroscopy data such as MUSE data. It was originally written to be used for PHANGS-MUSE and is described in \cite{Emsellem2022}. 

The DAP follows several steps. In a first step, the individual spectra are re-binned to a logarithmic wavelength sampling and corrected for foreground extinction using the extinction law from \cite{Cardelli1989} with a user-supplied $E(B-V)$ value. For NGC\,1427a, we selected $E(B-V) = 0.012$ according to the extinction map from \cite{Schlegel1998}. Then, the data are spatially binned with the two-dimensional Voronoi binning scheme \citep{Cappellari2003} to a user-specified target signal-to-noise ratio ($S/N$). Finally, the binned spectra are fitted with the penalised pixel-fitting routine \textsc{pPXF} \citep{Cappellari2004, Cappellari2017, Cappellari2023} to obtain the parameters of the line-of-sight velocity distribution and stellar population properties. The reported velocities are relative to the user-supplied systemic velocity, which we set to $v_\text{sys}=2028$ km s$^{-1}$ \citep{Bureau1996}.

It is possible to fit the stellar and gaseous contribution to a spectrum simultaneously with \textsc{pPXF}. For the latter, \textsc{pPXF} also returns emission line fluxes and velocities which can be used in a further analysis of the ionised gas properties. 

For NGC\,1427A, we used Voronoi-binned maps with a \mbox{$S/N$ = 100}, resulting in binned maps with 416 bins. This is a sufficient number to obtain a detailed view of the stellar kinematics and population parameters. As described below, for the gas kinematics we explored the binned and spaxel-by-spaxel results that give a much higher spatial sampling. For fitting with DAP, we use the wavelength range from 4700 \AA\, to 7050 \AA\, and mask sky line residuals. 

\subsection{Stellar and gas kinematics} 
\begin{figure}
    \centering
    \includegraphics[width=0.48\textwidth]{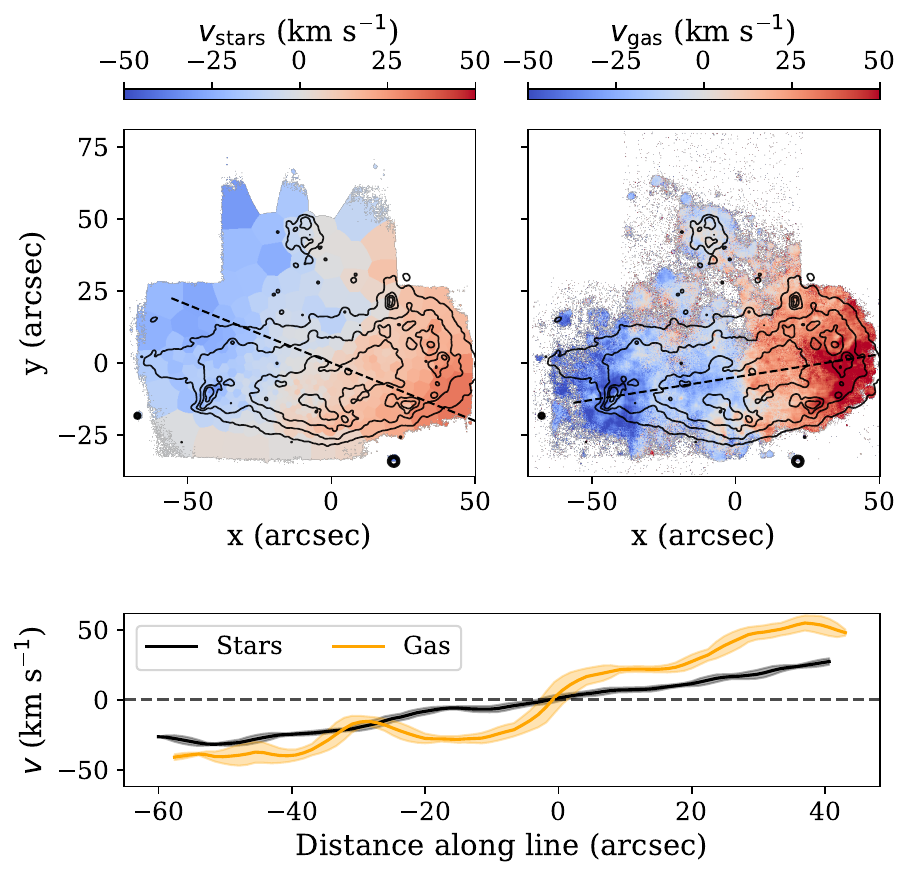}
    \caption{Stellar (top left) and H$\alpha$ (top right) velocity maps. The left panel shows a Voronoi-binned map with $S/N$ = 100, while the right panel shows the H$\alpha$ velocity from a spaxel-by-spaxel analysis. The black contours show the flux from the FDS $g$-band image ($\mu_g$ from 22 to 23.5 mag arcsec$^{-1}$ in steps of 0.5). The dashed lines indicate axis of maximum rotation, and the bottom panels shows the extracted mean velocity profiles along these lines $\pm$ 2 arcsec. Shaded areas indicate standard deviations.}
    \label{fig:MUSE_kinematics}
\end{figure}

Figure \ref{fig:MUSE_kinematics} shows the kinematics for the stellar and gaseous component. The latter is traced here by the H$\alpha$ line, but other emission lines give similar velocity fields. We notice that NGC\,1427A appears to be rotating in the stars and the ionised gas, but both components show a different velocity field. While the rotation axis of the gas appears to be perpendicular to the bar-like shape of NGC\,1427A, the axis of the stellar rotation might be inclined. \cite{Carvajal2026} measured a difference in the rotation angle of $\Delta\theta \sim 31^{\circ}$. The bottom panel of Fig. \ref{fig:MUSE_kinematics} shows the velocity values extracted along the axis of maximum rotation, using a 100\arcsec\,long and 6\arcsec\,wide box to robustly determine the mean and standard deviation. As seen from the maps and these profiles, the gas reaches higher rotation amplitudes of $\sim \pm$ 50 km s$^{-1}$, while the stellar rotation only reaches $\sim \pm$ 25 km s$^{-1}$. This is not surprising, since it is expected that the stars (especially older populations) are dynamically hotter with a larger velocity dispersion contributing to the overall motion, while the ionised gas tends to be rotationally supported (e.g. \citealt{Shetty2020}).  
We find stellar velocity dispersions of $\sim 20$ km s$^{-1}$ across the FOV, and gas velocity dispersions $< 15$ km s$^{-1}$. However, those values are much smaller than the typical instrumental resolution of MUSE ($\sim 80$ km s$^{-1}$), so the derived $\sigma$ per spaxel or bin is likely very uncertain. 

As also noted from the  \hi\ kinematics presented in \cite{Serra2024}, the gas kinematics within NGC\,1427A appear rather regular with no obvious sign of disturbances as could be expected from a recent merger. Also in the stellar kinematics, the northern object follows the rotation of the galaxy. 

\begin{figure*}
    \centering
    \includegraphics[width=0.80\textwidth]{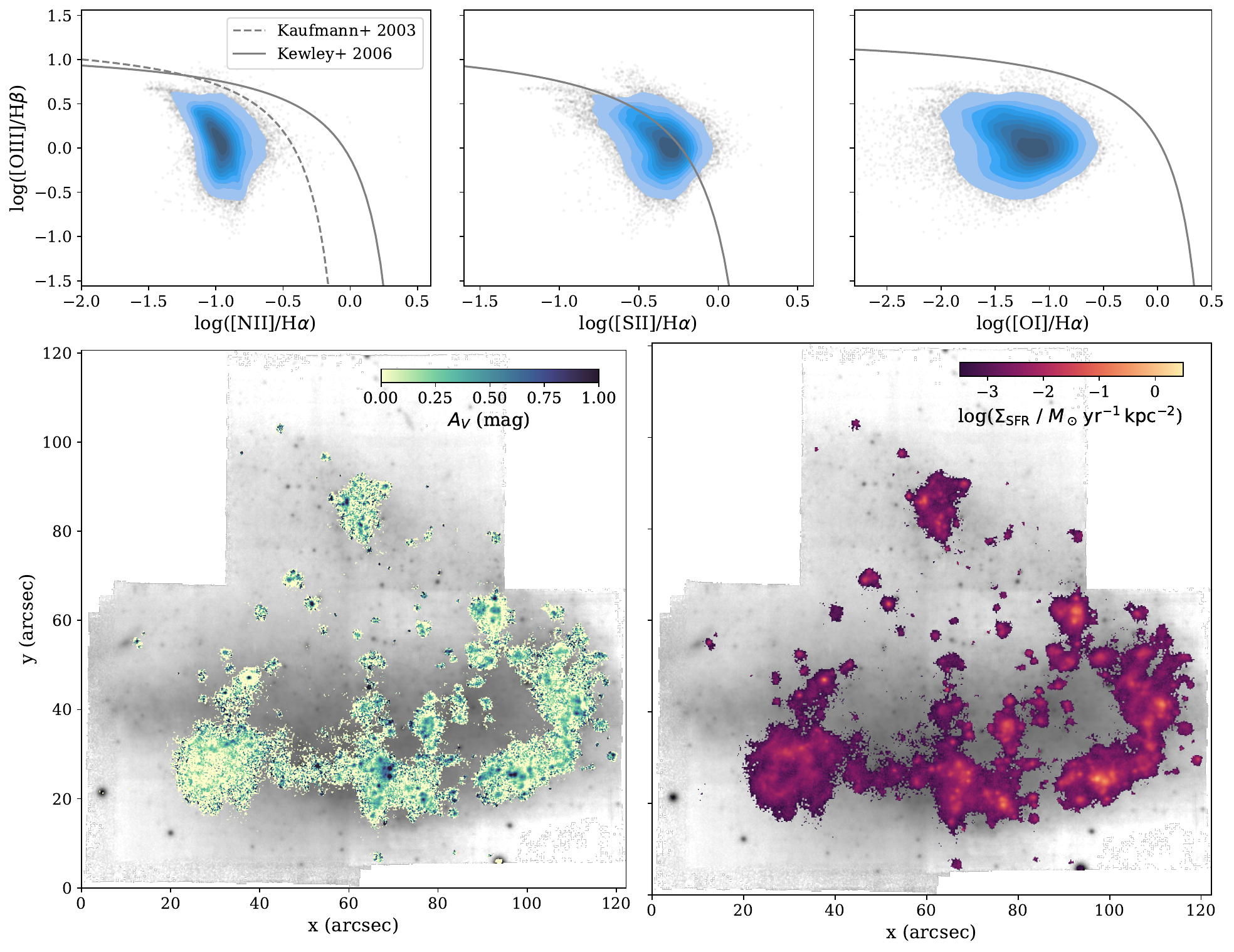}
    \caption{Overview of the emission line analysis. Top panels: BPT diagrams showing the flux ratios of [OIII]/H$\beta$ in comparison with [NII]$_\text{6584\AA}$/H$\alpha$ (left), [SII]$_\text{6717+6731\AA}$/H$\alpha$ (middle), and [OI]$_\text{6300\AA}$/H$\alpha$ (right). Only spaxels where the relevant lines have $S/N > 5$ are shown. The solid grey lines show the demarkation line between AGN and star formation regions from \cite{Kewley2006}, the grey dashed line shows the demarkation lines from \cite{Kaufmann2003}. Bottom panels: Extinction map (left) showing the V-band extinction assuming the extinction law from \cite{Calzetti2000}. Right: Surface density of the SFR obtained by converting the H$\alpha$ flux. The greyscale image in the background shows the flux of the MUSE white-light image for visualisation.}
    \label{fig:emission_lines}
\end{figure*}

\subsection{Gas analysis}
\label{sect:gas_analysis}
As one output, DAP returns emission line fluxes for a range of different emission lines including H$\beta$, H$\alpha$, [O\,{\sc iii}], [N\,{\sc ii}], and [S\,{\sc ii}]. From those, we can infer the star formation rate (SFR), dust attenuation, and gas metallicity. For a full analysis of the gas diagnostics, we refer to \cite{Carvajal2026} and follow-up work (Carvajal et al. prep). We focused on the extinction, which is used later to constrain extinction in the SED fitting (Sect.~\ref{sect:SED_fitting}), and the SFR to compare against the position of young clusters. In the following, we only consider spaxels that have a $S/N > 5$ for the used lines.

\subsubsection{Dust attenuation}
While DAP already corrects for foreground extinction, the intrinsic dust attenuation has to be accounted for. To do so, we use the Balmer decrement given by the flux ratio H$\alpha$/H$\beta$. This value has an intrinsic value of 2.86 (case B recombination,  \citealt{OsterbrockFerland2006}), and hence, measuring this ratio allows us to infer the colour excess $E(B-V)$ caused by intrinsic extinction (see \citealt{Calzetti2000}), 
\begin{equation}
    E(B-V) = 1.97\,\text{log}\left(\frac{\text{H}\alpha/\text{H}\beta}{2.86}\right).
\end{equation}

This can be used to map dust extinction across the MUSE FOV as Fig. \ref{fig:emission_lines} shows. Further, it allows us to infer the intrinsic H$\alpha$ luminosity,
\begin{equation}
    L_\text{int}(\text{H}\alpha) = 10^{0.4 A_{\text{H}\alpha}} L_\text{obs}, 
\end{equation}
with attenuation $A_{\text{H}\alpha} = k($H$\alpha) E(B-V)$, where $k({\rm H}\alpha)$ is the value of the extinction curve at H$\alpha$. We used the \cite{Calzetti2000} extinction curve for star bursting galaxies, with $R_V = A_V / E(B-V) = 4.05$, to correct the measured emission line fluxes for extinction. We chose this extinction law to ensure self-consistency with the SED modelling described in Sect. \ref{sect:SED_fitting}, where the \cite{Calzetti2000} is used as well. As can be seen from the map in the bottom left panel in Fig. \ref{fig:emission_lines}, the extinction is overall fairly low, but reaches $A_V > 1$ mag in a few regions of intense star formation. The low extinction is not particularly surprising since NGC\,1427A is a low-metallicity dwarf galaxy. 

\subsubsection{Ionisation sources from emission line ratios}
Before continuing with the analysis of the emission lines, we tested whether the emission is caused by star formation or rather from other ionisation sources more associated with shocks (e.g. active galactic nuclei). For this reason, we plot emission line fluxes in BPT diagrams in the top panels of Fig.~\ref{fig:emission_lines}. As can be seen, almost all spaxels fall well below the star formation demarcation lines from \cite{Kewley2001} when using the classical BPT diagram (flux ratios [N\,{\sc ii}]/H$_\alpha$ vs [O\,{\sc iii}]/H$_\beta$) or the [O\,{\sc i}]/H$_\alpha$ vs [O\,{\sc iii}]/H$_\beta$ diagram. In the [S\,{\sc ii}]/H$_\alpha$ vs [O\,{\sc iii}]/H$_\beta$ diagram, several spaxels are above the demarcation line. Checking the spatial distribution of those spaxels, we find them to be located at the edges of the star-forming regions, indicating that the emission in those regions likely becomes dominated by diffuse ionised gas or outflows from star formation (e.g. \citealt{Belfiore2022}). In summary, the emission seen in NGC\,1427A appears to be created by star formation. A follow-up paper (Carvajal et al. in prep.) will discuss the ionisation diagnostics in more details.

\subsubsection{Star formation rate density}
From the extinction-corrected H$\alpha$ fluxes in each of the star-forming spaxels, we calculate the H$\alpha$ luminosity assuming a distance of 20 Mpc \citep{Blakeslee2009} to the Fornax cluster. From this, the SFR density can be estimated using the relation from \citep{Hao2011},
\begin{equation}
    \text{SFR}(M_\odot \text{yr}^{-1}) = 10^{-41.27} L_\text{int}(\text{H}\alpha),
\end{equation}
with $L_\text{int}(\text{H}\alpha$) in erg s$^{-1}$. 

The right panel in Fig.~\ref{fig:emission_lines} shows the SFR surface density ($\Sigma_{\rm SFR}$) for NGC\,1427A in the MUSE FOV. As this map shows, the star formation is mainly located along the edges of the stellar body and in the northern object. Several distinct peaks are seen, likely indicating the position of star-forming regions and clumps.

We find a total SFR of \mbox{0.08\, $M_\sun$ yr$^{-1}$} in this MUSE FOV. In their analysis of the MUSE data, \cite{Carvajal2026} find a similar value of \mbox{SFR = 0.07\, $M_\sun$ yr$^{-1}$} from the H$\alpha$ flux. These two spectroscopic values are slightly higher than the values reported by \cite{Mora2015} of 0.05 $M_\sun$ yr$^{-1}$, and \cite{Georgiev2006}, who found 0.057 $M_\sun$ yr$^{-1}$, both based on H$\alpha$ narrow-band photometry. The differences are likely the result of the diverse extinction treatments and adopted distances.  

\begin{figure}
    \centering
    \includegraphics[width=0.48\textwidth]{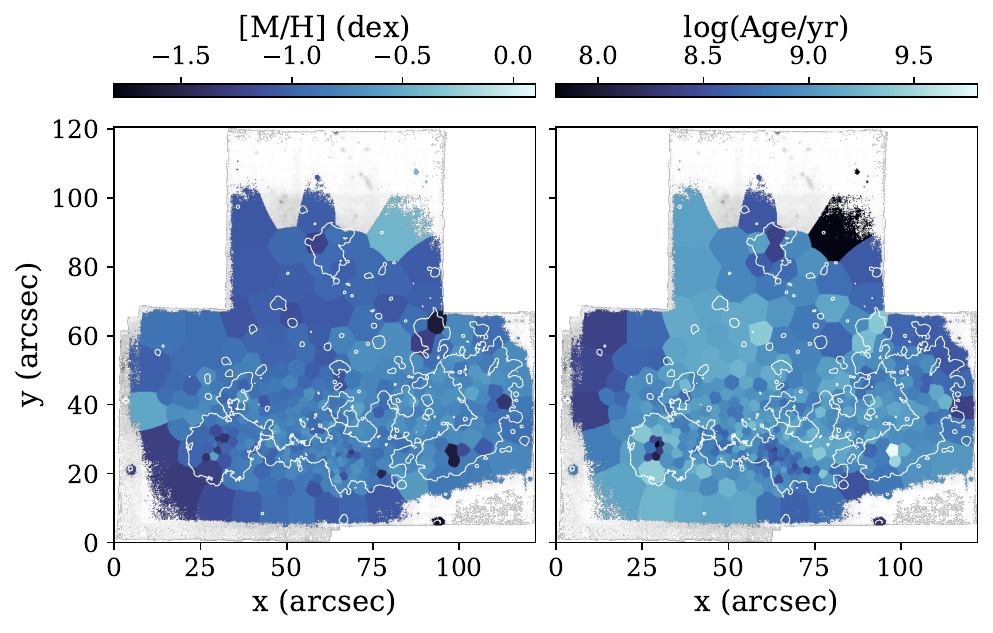}
    \caption{Maps of stellar population properties. Left: Light-weighted metallicity. Right: Light-weighted age. The greyscale image in the background shows the flux image of the MUSE white-light image. The white contours show the extent of the H$\alpha$ flux to guide the eye.}
    \label{fig:SSP_results}
\end{figure}

\subsection{Stellar ages and metallicities}
\label{sect:ages_and_metals}
To fit the stellar populations with DAP, a library of stellar population models has to be provided. We chose the E-MILES models\footnote{\url{miles.iac.es}} \citep{Vazdekis2010, Vazdekis2016}, which are well-established stellar population models based on empirical spectra. Typically, the E-MILES models based on the Padova+00 isochrones \citep{Giradi2000} cover ages from 63 Myr and 17.78 Gyr. Since NGC\,1427A is a star-forming dwarf galaxy that likely hosts younger stars than this lower age limit, we complemented the standard E-MILES library with the younger models described in \cite{Asad2017}, which reach ages of 6.3 Myr. To create a regular grid of models, we selected ages between 6.3 Myr and 15.8 Gyr and metallicities between [M/H] = $-$1.71 and 0.0 dex. In total, we used 295 different models.

We show the resulting light-weighted metallicity and age maps in Fig.~\ref{fig:SSP_results}. Only bins with $S/N > 90$ are shown. As can be seen, the maps are overall relatively flat with weak trends showing lower metallicities at older ages. The metallicity of the northern object (mean metallicity of \mbox{$\sim -1.0$ dex}) appears to be slightly lower than that of the main body ($\langle$[M/H]$\rangle$ $= -0.92$ dex), similar to the region between the main body and the northern object.

DAP also produces mass-weighted maps that show similar behaviour, however, with offset mean values. While the mean light-weighted age of NGC\,1427A across the FOV is $\sim 800$ Myr, the mean mass-weighted age is $\sim$ 6 Gyr, illustrating that even though NGC\,1427A is forming stars at the moment, most of the mass is still found within the old populations.

\section{Star cluster properties}
\label{sect:SED_fitting}
The main focus of this paper is to understand the evolutionary history of NGC\,1427A from the integrated light of the stellar body and its star cluster population. To analyse the star clusters, we used an SED fitting of star cluster candidates using \textsc{bagpipes}\footnote{\url{https://bagpipes.readthedocs.io/en/latest/}} \citep{Carnall2018, Carnall2019, Carnall2020}, a Bayesian method to fit SEDs with flexible models. 

\begin{figure}
    \centering
    \includegraphics[width=0.45\textwidth]{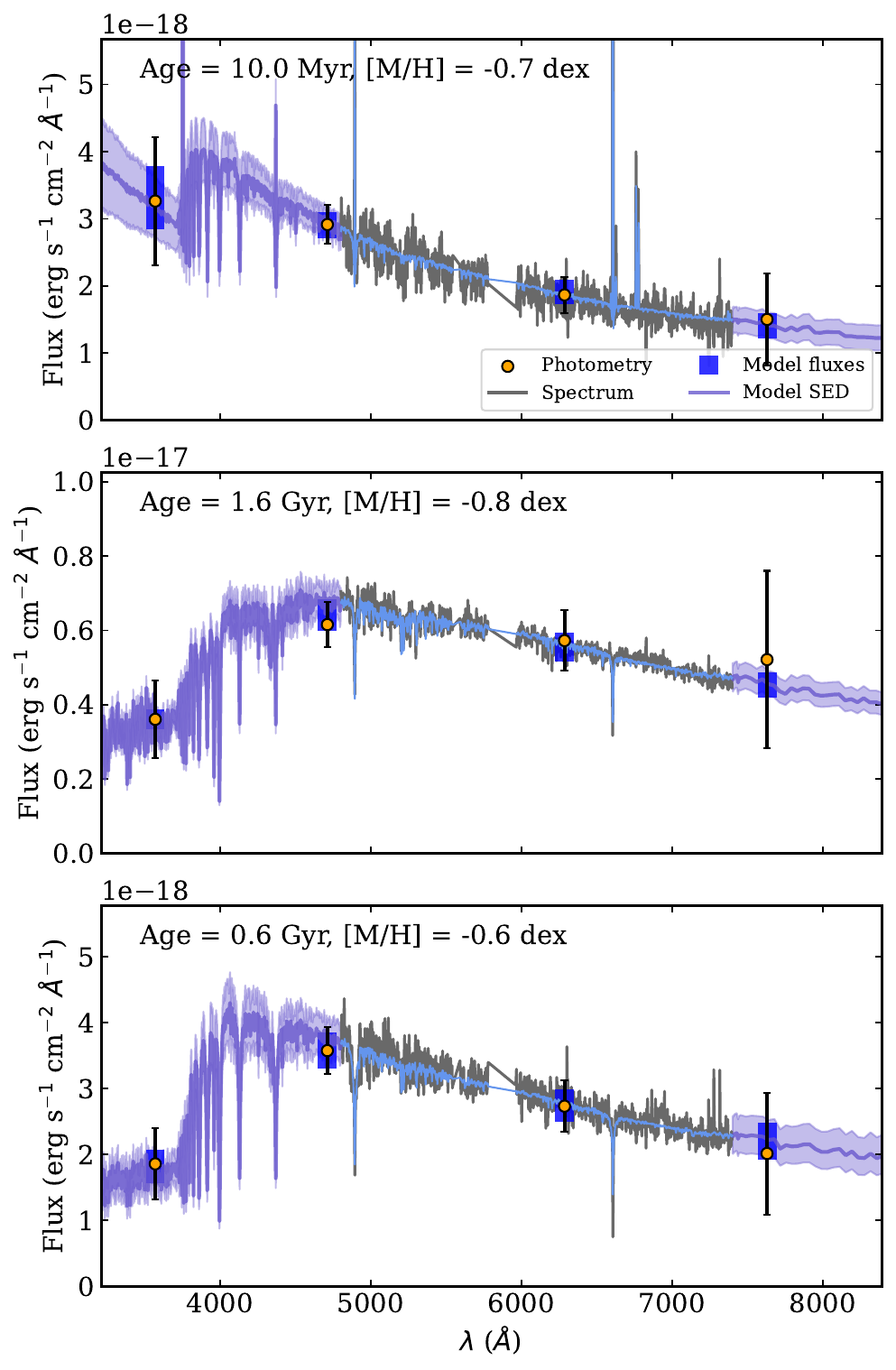}
    \caption{Examples of successful SED fits. For these star clusters, the FDS fluxes (orange) and the MUSE spectra (grey) were fitted with \textsc{bagpipes}. The best-fit model SED is shown in purple in the region without spectroscopy and in blue in the region of the MUSE spectrum. Corresponding model fluxes are shown by the dark blue squares. The shaded regions and the extension of the squares span between the 16th and 84th percentiles of the posteriors. Best-fitting age and metallicity are shown in the top left panel.}
    \label{fig:SED_fits}
\end{figure}

\begin{figure}
    \centering
    \includegraphics[width=0.45\textwidth]{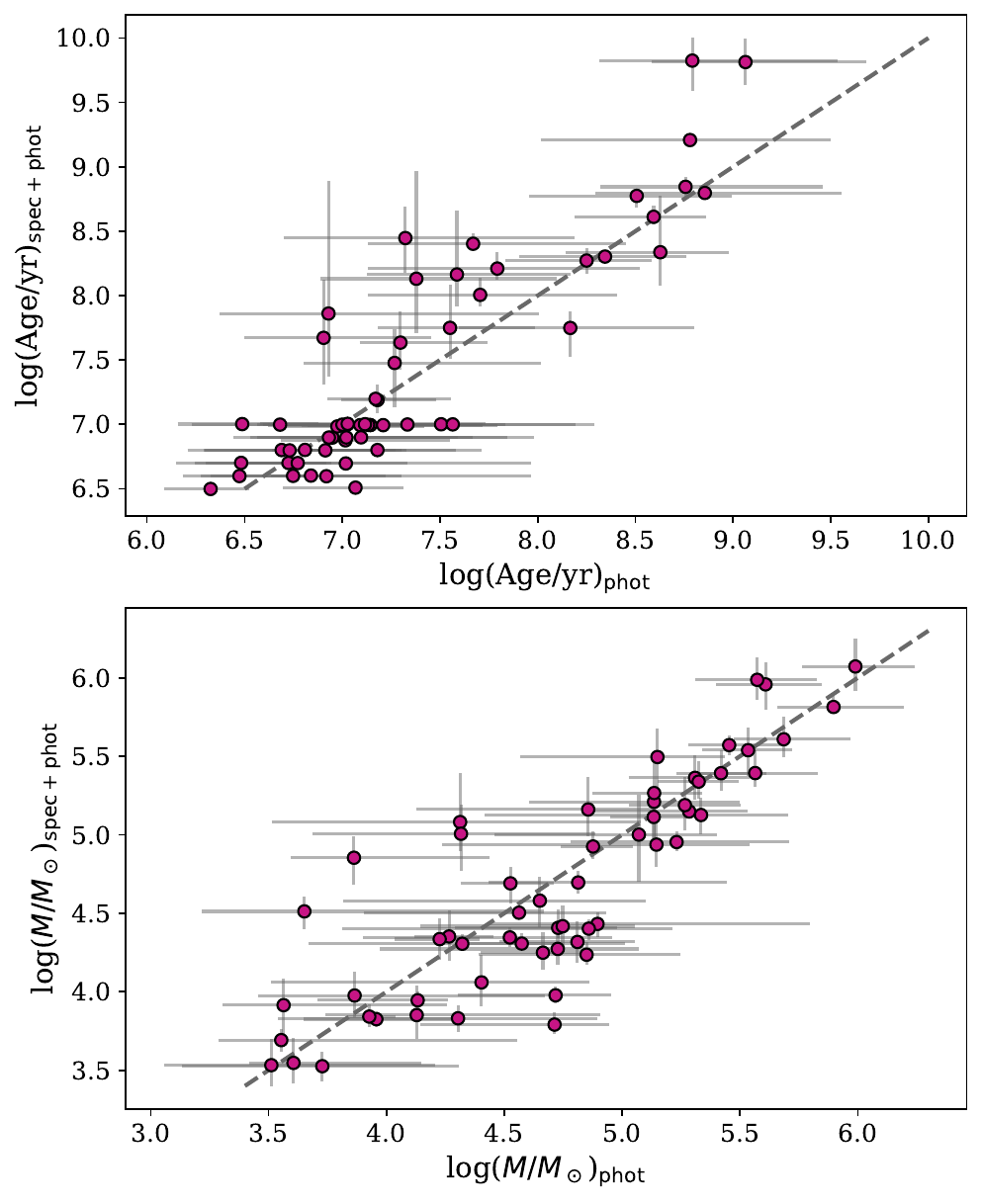}
    \caption{Comparison between SED fit results for photometric measurements alone and combined with spectroscopy. Top: Best-fit age. Bottom: Best-fit mass.}
    \label{fig:comparison_SED_fits}
\end{figure}

\subsection{Fitting photometric candidates}
Typically, SEDs are fit using photometry in multiple bands to obtain redshifts, masses, and SFHs of galaxies in the local Universe and at higher redshifts (e.g. \citealt{Carnall2024, Leung2025}). We did not wish to constrain the SFH of the full galaxy with SED fitting, but rather the ages and masses of the star clusters to employ them as surviving tracers of main star formation bursts. For this reason, we started by running \textsc{bagpipes} on all sources in the FDS catalogue using the FDS magnitudes in the $u, g, r,$ and $i$ filters with their reported uncertainties. 

The \textsc{bagpipes} code is implemented in Python and provides flexible SED fitting to describe the stars, gas, and dust of galaxy spectra. The user can build models of varying complexity by including different prescriptions for the SFH, dust absorption, or nebular emission. Within \textsc{bagpipes}, the \cite{BruzualCharlot2003} stellar population models (2016 version) are included with a Kroupa initial mass function \citep{Kroupa2001}. Different prescriptions for dust attenuation are available, including the extinction law from \cite{Calzetti2000}. The nebular emission is modelled using the \textsc{CLOUDY} code \citep{Ferland2017}.

To limit the complexity of the fitted model, we used a single burst model to describe the SFH of the star clusters that is defined by a single age, metallicity, mass, and extinction. We included nebular emission with a fixed ionisation parameter (set to \mbox{log$(U)=-3$}\footnote{We also tried other values for log$(U)$ and found no significant change to the other parameters, especially in these photometric fits.}) that adds emission lines for populations younger than $\sim$ 10 Myr, and we fixed the redshift to that of NGC\,1427a. Thanks to the MUSE data, we can already constrain the extinction to be rather low. For all star cluster candidates where the extinction map is greater than $0$, we used Gaussian priors on $A_V$ from MUSE with a dispersion of $\sigma$ = 0.2 mag. In other cases, we allow conservative flat priors between 0 and 2 mag. Due to the flat overall metallicity profile, we fixed the metallicity to $-0.5$ dex when only fitting the photometry. Doing so, only age and mass are fully free in the fitting, which alleviates the disadvantage of only using four filters.

\subsection{Spectroscopic candidates}
\label{sect:bagpipes_spectra}
For a subsample of bright clusters, we used \textsc{bagpipes} to fit the MUSE spectra and the OmegaCam photometry simultaneously. The following describes how we extracted the spectra from MUSE before fitting with \textsc{bagpipes}.

\subsubsection{Extracting star cluster spectra}
To detect star cluster candidates in the MUSE data, we first subtracted the background from the MUSE white-light image using \textsc{photutils}\footnote{\url{https://photutils.readthedocs.io/en/stable/}} background estimation routines with a median filter of box size of 5 $\times$ 5 pixels and kernel size of 3 $\times$ 3 pixels. In this residual image, we detected all sources that are 3$\sigma$ above the background rms.
Matching those against sources that have magnitudes from the FDS catalogue yields 162 initial cluster candidates. 

For those, we extracted their spectra using a Gaussian-weighted circular aperture with a FWHM of 0.8\arcsec (4 pixels) to boost the signal from the compact sources. To account for the background, similar to aperture photometry, we also extracted a background spectrum in an annulus around each source using an inner radius of 5 pixel (1.0\arcsec) and outer radius of 7 pixels, and subtracted this background spectrum from the source spectrum. 

To match the flux with the photometry before SED fitting, we extracted synthetic $r$-band magnitudes from the MUSE spectra using \textsc{synphot} \citep{synphot} and multiplied the MUSE spectra with the ratio between the OmegaCam and synthetic magnitudes.

\begin{figure*}
    \centering
    \includegraphics[width=0.98\textwidth]{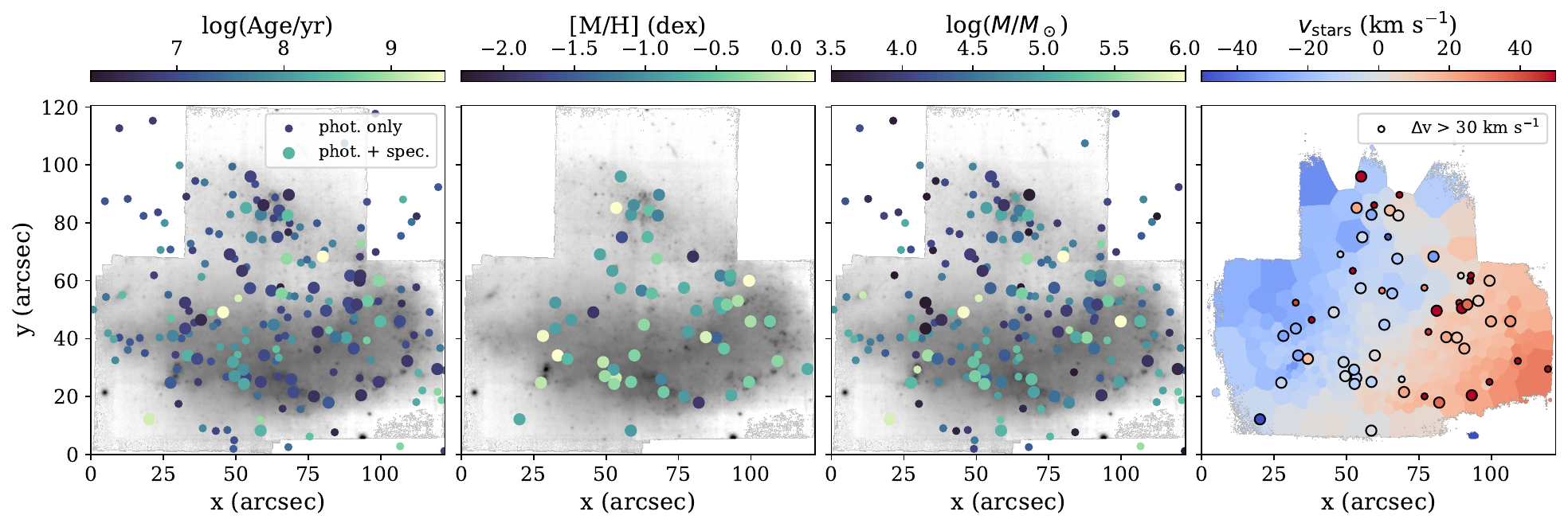}
    \caption{Results of the star cluster SED fitting overplotted on the image of NGC\,1427A. The first three panels show from left to right the star cluster ages, metallicities, and masses. Clusters where spectra and photometry were fitted simultaneously are shown as large symbols. Small symbols refer to photometric-only SED fits. Right panel: Binned stellar line-of-sight velocity field of NGC\,1427A with star cluster velocities overplotted as circles. Small circles refer to clusters with velocity uncertainties larger than 30 km s$^{-1}$.}
    \label{fig:cluster_on_image}
\end{figure*}

\subsubsection{Simultaneous fit of photometry and spectroscopy}
In addition to the photometric fits, we used \textsc{bagpipes} to fit the cluster candidates that have FDS photometry and a MUSE spectrum. \textsc{bagpipes} offers the possibility to fit photometry and spectroscopy simultaneously, as the stellar population models from \cite{BruzualCharlot2003} have been updated to provide model spectra at a resolution of 2.5 \AA\, from 3525 to 7500 \AA\, using the MILES stellar spectral library \citep{FalconBarroso2011}. 
In the spectroscopic fits we also used a single burst model, and again constrained the extinction based on the map from Fig. \ref{fig:emission_lines}, but we left the metallicities free. 

We inspected the fits manually and found that the fit fails in some cases, especially when the signal-to-noise of the spectrum is low ($< 5$ pix$^{-1}$ in the continuum). In other cases, the photometry does not seem to follow any SED within the model library, often because the $u$-band flux appears to be much higher than seems reasonable. To get a reliable sample, we removed the cluster candidates in which no convincing fit to the spectroscopy and photometry was found, as well as the ones where even the photometry-only fits were not able to reproduce the fluxes. In the end, we keep 58 star clusters with good fits to spectroscopy and photometry.
In Fig. \ref{fig:SED_fits} we show three examples of successful fits. For comparison, Fig. \ref{fig:bad_SED_fits} shows three examples of unsuccessful fits.

\subsection{Comparing photometric and spectroscopic fits}
\label{sect:comparison_SEDs}

The 58 clusters with good quality SED fits to photometry and spectroscopy allowed us to test how reliable the results become when only the photometry based on the four FDS filters is fitted. 
Figure \ref{fig:comparison_SED_fits} shows the comparison in derived cluster ages and masses between fitting photometry only and combining photometry and spectroscopy. As can be seen, the uncertainties are much smaller in the spectroscopic fits, but overall the results are comparable. Given that we only fit a very simple single burst model and could constrain the extinction from the MUSE data directly, this is perhaps not too surprising, but it is encouraging for the larger sample of clusters with only photometric measurements. 

For this reason, we selected our final star cluster catalogue using the 58 sources for which \textsc{bagpipes} was able to fit spectroscopy and photometry and complemented this sample with the cluster candidates without spectroscopy or low resolution spectra, but where the \textsc{bagpipes} models could reproduce the fluxes in the four FDS bands sufficiently. In total, this catalogue contains 222 sources. Besides the brightest two sources that could be identified as foreground stars, and two background galaxies that were visually identified, the selected sources cover almost the full magnitude range of the OmegaCam sample ($g = 25 - 20$ mag). 

\subsection{Star cluster properties across NGC\,1427A}

Figure \ref{fig:cluster_on_image} shows the results from the SED fitting. The first three panels show the ages, metallicities, and masses of the star clusters for the photometric-only and photometry + spectroscopy samples. While young clusters are distributed all over the galaxy, the brighter young clusters with available spectroscopy are mainly found along the edges of the galaxy and in the northern object where the SFRs reach higher values (Fig. \ref{fig:emission_lines}). In the inner regions of the galaxy, the star clusters are slightly older, and at least two old star clusters (globular clusters) are located within the galaxy, at least in projection. The metallicities of the star clusters varies, but no clear trend is discernable. The star cluster masses cover the range $\log(M/M_\odot)\approx3.5-6.0$, and seem to reach $\log(M/M_\odot)\approx5.-5.5$ at the edges of the galaxy whereas the star clusters in the northern object are less massive. 

The right panel of Fig.~\ref{fig:cluster_on_image} shows the line-of-sight velocities of the star clusters with available MUSE spectra. As the redshift was used as a fixed parameter in the \textsc{bagpipes} fits, those velocities were obtained by fitting the star cluster spectra directly with \textsc{pPXF}. We find that all 58 star clusters with spectroscopic data have stellar and gaseous velocities in agreement with NGC\,1427A, in a range between 1950 and 2100 km s$^{-1}$. In a few cases, especially for young clusters dominated by emission lines, the stellar velocities have uncertainties $\Delta v > 30$ km s$^{-1}$. As seen when placed on the stellar velocity map, the star clusters generally follow the rotation of the galaxy. Especially the older clusters closely follow the stellar velocity field of NGC\,1427A, while others have line-of-sight velocities that more closely match the gas velocities (Fig. \ref{fig:MUSE_kinematics}). A few clusters (especially ones with large velocity uncertainties) show slight deviations from the underlying rotation.   

\begin{figure}
    \centering
    \includegraphics[width=0.43\textwidth]{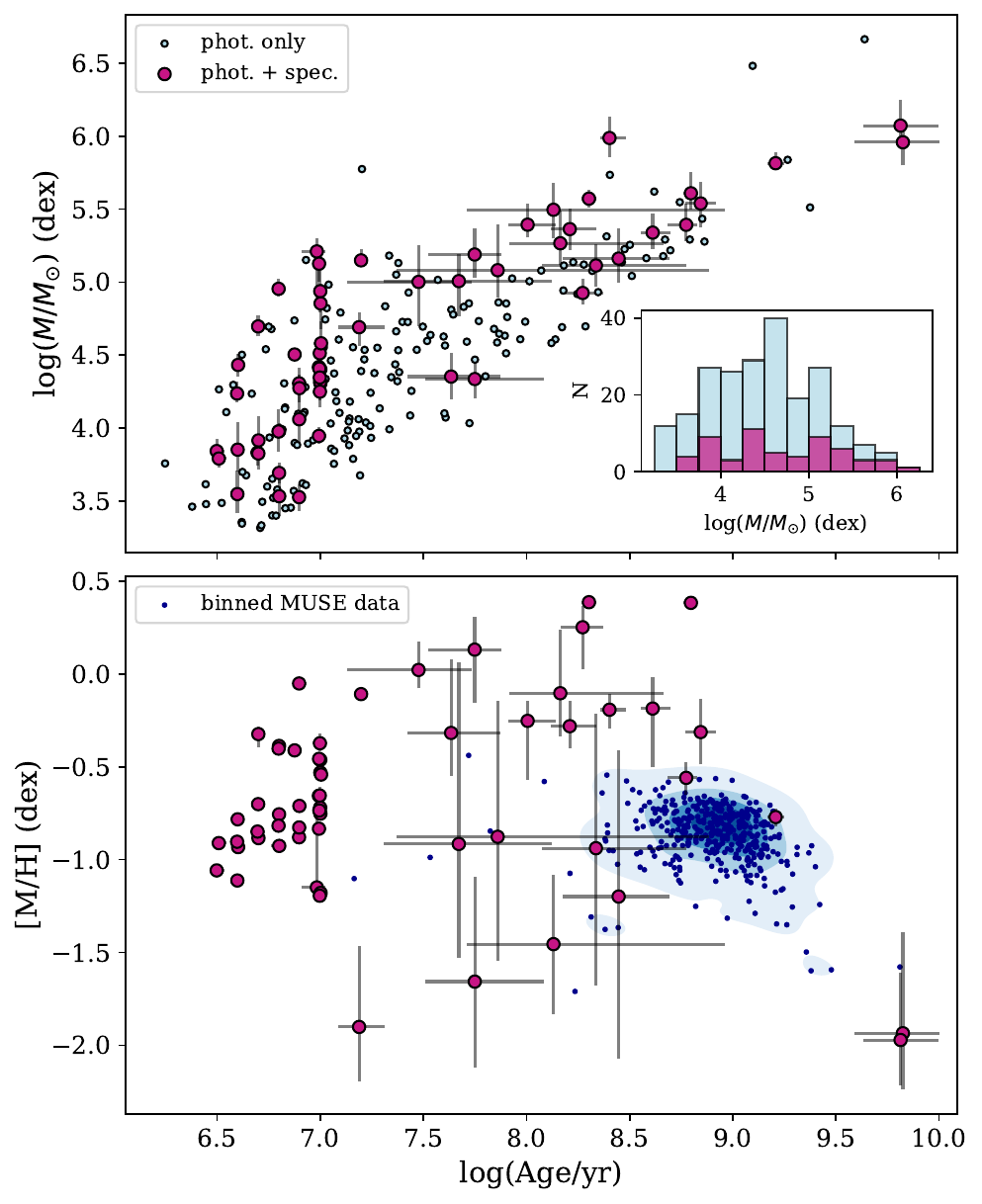}
    \caption{Age, masses, and metallicities of star clusters. Top: Age vs. mass, with photometric candidates in light blue. The inset shows histograms of the cluster masses. Bottom: Age vs. metallicity, only available for clusters with photometry and spectroscopy. Small blue points show the values from the binned maps (see Fig. \ref{fig:SSP_results}) and the blue shaded region shows their distribution with a kernel density estimator for visualisation.}
    \label{fig:age_metallicity_mass}
\end{figure}

Figure \ref{fig:age_metallicity_mass} shows the age-mass and age-metallicity distributions of the star clusters for which spectra and photometry were used in the SED fits. The age-mass plane shows a clear correlation in the sense that older clusters are more massive. We believe that this is caused by two effects: on the one hand, young clusters of the same stellar mass are much brighter in optical bands than their older counterparts, creating an age-dependent detection limit. On the other hand, low-mass clusters dissolve quickly in the tidal fields of galaxies (e.g. \citealt{Lamers2005, Gieles2006, Bastian2012}), and only massive clusters can survive for several hundreds of million years or even billions of years. Consequently, we also observe a larger scatter in stellar masses for younger clusters than for older ones. In any case, we find that no star clusters with masses $\log(M/M_\odot)\gtrsim5.5$ were formed in the last 100 Myr.

In the age--metallicity plane, the younger clusters (Age $< 10^7$ yr) show metallicities between $\sim -1.2$ to $-0.3$ dex, while the scatter for intermediate-age clusters is significantly larger, ranging from $\sim -2$ dex to solar metallicities. We believe that this scatter stems from a combination of factors. Several of the metal-poor, intermediate age clusters also have rather low $S/N$ < 10, which is reflected in large age and metallicity uncertainties. Additionally, the absence of emission lines in these clusters leads to larger uncertainties in the recovered ages. Better constraints on the properties of these clusters would come from information at longer or shorter wavelengths. The metallicity distribution of the young clusters compares well with the light-weighted mean metallicity from the stellar population fits ($\sim -1$ dex). 
The two oldest clusters are also some of the most massive and most metal-poor ones ($-2.0$ dex) and resemble Milky Way halo globular clusters. The age-metallicity relation of the old star cluster population with $\log{(\rm Age/yr)}\gtrsim8.5$ suggests a gradual chemical enrichment of NGC\,1427A with time. 

\begin{figure*}
    \centering
    \includegraphics[width=0.95\textwidth]{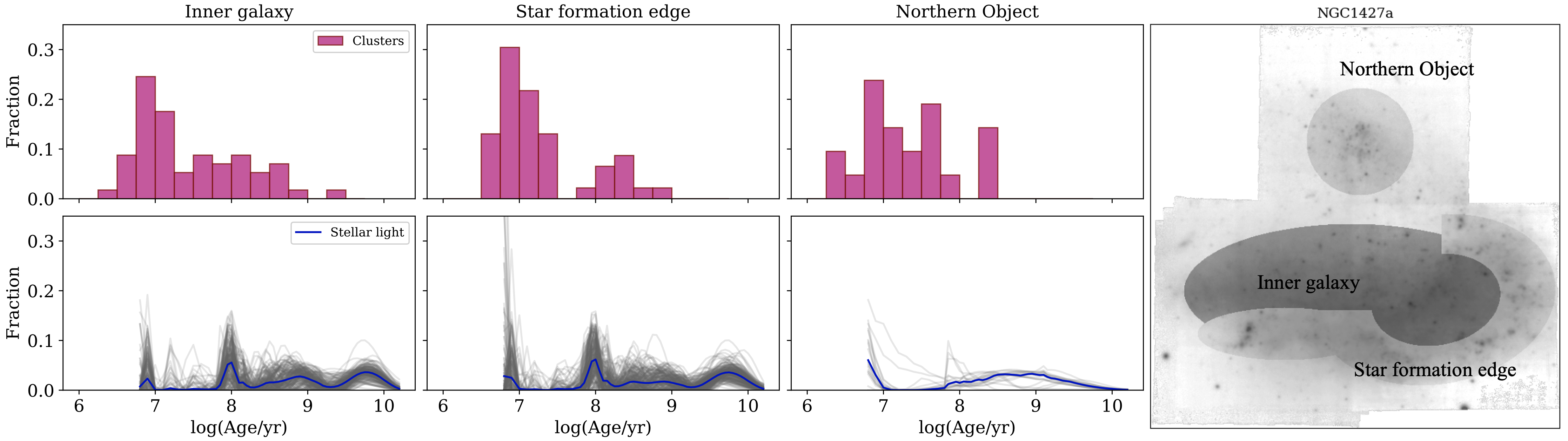}
    \caption{Star formation history of NGC\,1427A as traced by the star clusters (pink histograms) and from the integrated stellar light (grey and blue lines). The different panels show the distributions for different regions in the galaxy as indicated by the different shading in the rightmost panel. In each panel, the SFHs for the different bins from the DAP analysis are shown in the grey lines, while the blue lines refer to the mean.}
    \label{fig:SFH_comparison}
\end{figure*}

\section{The star formation history of NGC 1427A}
\label{sect:SFH}
In this section, we describe the SFH of NGC\,1427A as revealed by the MUSE data and its star clusters. We then set this in context with the results from other works to describe the evolutionary history of NGC\,1427A.

\subsection{Comparing star clusters and stars}
The rich data set for NGC\,1427A allows us to study the SFH of NGC\,1427A in two ways. Firstly, full spectrum fitting with DAP as a wrapper around \textsc{pPXF} can be used to obtain the best fitting distribution of ages and metallicities that reproduce the input spectra. Based on the regularised weights returned by \textsc{pPXF}, mean ages can be recovered, but also the SFH as mass- or light fraction versus age can be extracted. This approach has been used to study the SFHs of early- and late-type galaxies and their stellar components (e.g. \citealt{Pinna2019, deSaFreitas2025}). 
From the fits to the binned data, such non-parametric SFHs are available for all bins in the MUSE data.

Alternatively, we can use the star cluster ages as probes of the NGC\,1427A SFH under the assumption that they were formed during intense star formation episodes.
To compare the two approaches, we show the cluster age distributions in different regions as well as the corresponding light-weighted SFHs from \textsc{pPXF} from the binned MUSE data in Fig. \ref{fig:SFH_comparison} as histograms and grey curves, respectively. We differentiated here between the star-forming edge of the galaxy, the inner galaxy, and the northern object. 
The figure shows reasonable agreement between the SFH traced by the star clusters and from the full spectrum fitting of the stellar body. In both cases (i.e. clusters and stellar body), multiple peaks are visible and trends between regions become evident. While the inner galaxy and the star-forming edge show both an extended SFH and a cluster age distribution with multiple peaks, the star-forming edge shows a higher fraction of young clusters with $\log({\rm Age/yr)}\lesssim7.5$ and a larger scatter towards young ages in the SFH derived from \textsc{pPXF}. Nonetheless, the star-forming edge and the inner galaxy show peaks in the SFH and the cluster age distributions for ages $\sim\!100$\,Myr, with a slightly broader peak for the inner galaxy region. Additionally, both regions show a broad peak at old ages ($>$ 5 Gyr), suggesting an extended underlying distribution of old stellar populations.

Within the northern object, star clusters show a relatively broad age distributions from a few million years to several hundred million years, but the total number of clusters is small. The \textsc{pPXF}-derived SFH shows a unique distribution compared to the rest of the galaxy, with a broad component at old ages and a peak at the youngest ages. As the stellar population templates only contain ages down to 6.3 Myr, it is possible that even younger ages might be present. Nonetheless, old populations in the SFH as well as the broad cluster age distribution suggest (together with the comparable kinematics) that the northern object is simply a regular region of NGC\,1427A that appears as a distinct environment only due to its recent star formation activity.

\subsection{Possible caveats}
The comparison between star cluster ages and the SFH derived from the integrated spectra could identify common trends, which is encouraging in building confidence in the results. Yet, while cluster ages and masses are likely reliable on a population level, individual measurements might be affected by biases. 

Especially, the photometry-only sample can still be biased due to the low number of filters used in the fitting. For example, it is possible that this sample still contains a few contaminant sources such as faint foreground stars or unresolved background galaxies. Additionally, a broader wavelength coverage towards the infrared could help to further constrain the ages and detect still embedded clusters (e.g. \citealt{Rodriguez2025}). We tested matching with available VIRCam data from the Vista Hemisphere Survey \citep{VHS}, but unfortunately these infrared observations are too shallow. 

Additionally, the SFHs derived from \textsc{pPXF} within \textsc{DAP} also have systematic uncertainties. As previously mentioned, the stellar population models with the youngest ages reach 6.3 Myr, while younger populations might be present in the galaxy. Moreover, the nebular emission is not modelled self-consistently within \textsc{pPXF} to constrain the ages. Instead, the SFHs are derived from the stellar components after removing the emission lines.

\section{The past of NGC\,1427A}
\label{sect:discussion}
Starting from the early ground-based observations presented by \cite{CelloneForte1997}, the formation and evolutionary history of NGC\,1427A has been debated. The findings made in the last two decades, have now started to paint a coherent picture of the past of NGC\,1427A within the Fornax cluster environment. 

\subsection{Recent star formation activity}
Any optical imaging of NGC\,1427A immediately reveals it to be a star-forming galaxy with a peculiar morphology and a wealth of star-forming regions (e.g. \citealt{CelloneForte1997}). \cite{Hilker1997} presented multi-band imaging of NGC\,1427A and identified numerous young star clusters, especially along the southern edge. They originally estimated ages $\leq$ 2 Gyr, while \cite{Mora2015} estimated that the most recent star formation activity took place $\sim$ 4 Myr ago. Our results are consistent with both studies, but now give a more detailed insight into the SFH of NGC\,1427A.

As Fig.~\ref{fig:SFH_comparison} reveals, the cluster age distribution and the SFH from the analysis of the MUSE spectra show multiple peaks. The majority of our clusters have ages $< 1$ Gyr, with two peaks at $\sim$ 10 and $\sim$ 30 - 300 Myr, that are matched to peaks in the SFH. We find that some youngest clusters are found along the star-forming edge (see also \citealt{Georgiev2006}), as well as in the northern object, but especially the young clusters from the photometric sample are located over the full field.

The two peaks at $\sim$ 10 Myr and a few 100 Myr can be interpreted within the context of the passage of NGC\,1427A through the Fornax cluster. The presence of young clusters was originally interpreted as a consequence of ram-pressure stripping (e.g. \citealt{Hilker1997, Sivanandam2014, Mora2015}). However, \cite{LeeWaddell2018} identified an  \hi\ tail almost perpendicular to the star-forming edge, pointing towards the south-east (away from the centre of Fornax). They suggested that the star formation was instead caused by a recent tidal interaction $\sim$ 4 Myr ago. With their deeper MeerKAT observations, \cite{Serra2024} argued against ram-pressure stripping acting on the star-forming edge as no compression of \hi\ was observed on the south-west side of the galaxy. Instead, they proposed that a galaxy interaction and ram-pressure stripping together caused the star formation in NGC\,1427A and the extended  \hi\ tail towards the south-west. In their scenario, NGC\,1427A experienced a tidal interaction or merger with another Fornax galaxy that started around 300 Myr ago that has originally removed gas from the galaxy disc. Then, ram-pressure stripping has further accelerated this gas and shaped it into the star-less  \hi\ tail seen today. With a multi-wavelength analysis, \cite{Carvajal2026} refined the picture of ram-pressure stripping in NGC\,1427a further. They argue that the ram-pressure stripping is mainly acting along the line-of-sight, and reaches the ISM of NGC\,1427A, affecting the star formation processes within the galaxy.

Our finding of a population of star clusters with ages of a few hundred million years and the corresponding SFH peak is consistent with the galaxy interaction or merger as proposed by \cite{Serra2024}.  We also find a population of younger star clusters ($\sim$ 10 Myr), suggesting that this interaction is still ongoing or that another galaxy or tidal encounter has led to the most recent star formation episode in NGC\,1427A. While the broad spatial distribution of young clusters from the photometric sample could indicate a population of star clusters formed in stripped gas, further data on these clusters would be needed to confirm such a population. In the kinematics of star clusters with spectroscopy (Fig. \ref{fig:cluster_on_image}), we do not see evidence of clearly stripped clusters as they should have blue-shifted velocities with respect to the galaxy, similar to the blue-shifted \hi\ tail \cite{Serra2024}. 

\subsection{Older populations}
While the star cluster population mainly contains young clusters ($<$ 1 Gyr), the SFH reveals a prominent population of older stars ($>$ 1 Gyr), suggesting that NGC\,1427A has formed stars over many billion years. The presence of old stars was already found in early imaging based on the presence of a diffuse stellar background and redder colours beyond the star-forming edge (e.g. \citealt{CelloneForte1997, Hilker1997}). With the spatially resolved spectroscopy provided by MUSE, we found that this old population is present throughout NGC\,1427A. 

Further evidence for old populations comes from the presence of globular clusters, as identified by \cite{Georgiev2006}. In our sample, we also find two clusters with globular cluster-like ages ($>$ 6 Gyr) and masses ($\log[M/M_\odot]\approx6.0$), which are also metal-poor and found outside the star-forming edge. In general, \cite{Georgiev2006} found a rather smooth distribution of globular clusters, possibly reflecting the smoother distribution of the underlying old population. We also compared our sample to the candidates from \cite{Georgiev2006}, finding two matches with available spectroscopy. In both cases, the spectra show emission lines and the SED fits reported ages of a few tens of million years. Unfortunately, the remaining GCs from \cite{Georgiev2006} are either outside of the MUSE data footprint or too faint to recover a useful spectrum.

\subsection{The northern object}
Originally proposed as the sign of an ongoing merger \citep{CelloneForte1997}, long-slit data presented by \cite{Chaname2000} found that the ionised gas kinematics of the northern object are consistent with the main galaxy. Similarly, \cite{Serra2024} found a smooth  \hi\ velocity field, indicating that this object is either a regular region within the galaxy or that any clear kinematic signature of a past merger has already been erased. With MUSE, we could confirm the coherent velocity field in the gas, but could also study the stellar kinematics. While there seems to be a general misalignment in the stellar and gas rotation axes (see \citealt{Carvajal2026}), the stellar kinematics of the northern object are consistent with the rest of the galaxy and no obvious disturbances that could be caused by a recent merger are evident.

Regardless of the kinematics, the stellar populations of the northern object do not set it apart from the main body of NGC\,1427A. While perhaps slightly more metal-poor than NGC\,1427A on average, the northern object does not appear as distinct component when studying the maps of ages, metallicities, or SFR. Similarly, the star clusters found in the northern object have similar ages as the young clusters found along the star formation edge and the SFH obtained from the integrated spectra reveal a similar underlying old population as in the rest of the galaxy. Taken together, all these results indicate that the northern object is an ordinary component of NGC\,1427A that appears as a distinct region due to its recent star formation activity. However, while the northern object itself may not be the remnant of a recent merger, it is possible that its star formation activity was triggered due to the same recent interactions or tidal forces that have also triggered star formation in the other regions of NGC\,1427A \citep{Serra2024}.

\section{Conclusions}
\label{sect:conclusion}
We presented a spectroscopic and photometric analysis of NGC\,1427A, a low-mass star-forming galaxy in the Fornax cluster with a peculiar morphology likely caused by its passage through the Fornax galaxy cluster. In particular, we focused on the star cluster content and SFH of NGC\,1427A. Our main conclusions are summarised below.
\begin{itemize}
    \item The velocity field of NGC\,1427A based on MUSE data reveals rather regular rotation, in a similar fashion as indicated by larger-scale \hi\ kinematics. However, the gas and stellar velocity fields appear to be misaligned, and the gas velocities reach higher amplitudes than the stars, suggesting that the stellar component is more dominated by random motions than the gas. The northern object does not appear as a kinematically distinct component, suggesting that it is part of the disc of NGC\,1427A. 
    \item The peculiar morphology of NGC\,1427A is driven by ongoing star formation and emission lines in the outskirts of the galaxy, which form a ring-like structure. In addition, the galaxy has an older main body. The northern object appears to be forming stars, but it is otherwise not distinct from the rest of the galaxy.
    \item Stellar population analysis from the MUSE data found overall flat distributions of stellar ages and metallicities. Overall, the galaxy appears to be metal poor ($\sim -0.9$ dex), with a light-weighted mean stellar age of $\sim$ 800 Myr, according to full spectrum fitting methods.  We note that these results can only account for populations older than 6.3 Myr, but the presence of H$\alpha$ indicates populations younger than $\sim$ 10 Myr. In contrast, the mass-weighted mean age is $\sim$ 6 Gyr, illustrating the mass-dominant old populations within NGC\,1427a.
    \item We applied SED-fitting methods to derive ages and masses of 222 star cluster candidates identified from ground-based OmegaCam photometry. For a subsample of 58 clusters, MUSE spectroscopy was integrated in the SED fitting. The comparison of spectroscopic and photometric fits found a relatively good agreement in ages and masses. 
    \item The star clusters show ages between $\sim 3$ Myr and several billion years, with masses down to $3 \times 10^{3}$ $M_\sun$. Two of the most massive clusters ($\sim 10^6 M_\sun$) are also the oldest, similar to metal-poor Milky Way globular clusters, with a trend of decreasing cluster mass with decreasing cluster age. The youngest clusters have the lowest masses, while intermediate-age clusters (a few 100 Myr to $\sim$ 1 Gyr) span a wider mass range. A handful of these clusters reach masses $> 3 \times 10^{5.5} M_\sun$ and might therefore be considered globular cluster progenitors.  
    \item The comparison of the SFH of NGC\,1427A based on a full spectrum fitting with the star cluster ages agreed in general, but highlighted that many of the star clusters represent the youngest population. We found a population of young clusters ($\sim$ 10 Myr) located along the south-western star-forming edge and in the northern object, as well as a population of intermediate-age clusters (a few hundred million years). The two populations have corresponding peaks in the SFH derived from fitting the MUSE spectra, which also shows an underlying old population ($>$ 1 Gyr) throughout NGC\,1427A.
    \item The evolutionary history of NGC\,1427A appears to have been complex. The old stars and its globular cluster population suggest that this galaxy has been forming stars over a prolonged time, and its SFH was significantly affected by the galaxy passage through the Fornax cluster. The intermediate-aged cluster population was possibly formed when NGC\,1427A interacted or merged with a galaxy a few hundred million years ago, as also suggested from \hi\ observations \cite{Serra2024}. However, the ionised gas emission and the young clusters further indicate recent and ongoing star formation, possibly triggered by a continuing interaction within Fornax.
\end{itemize}

We demonstrated once again that massive star clusters are lasting fossil tracers of star formation. Combined with other probes, they provide a promising path toward reconstructing the assembly history of a galaxy.

\begin{acknowledgements} We thank the referee for helpful suggestions that have improved this work.
KF acknowledges funding from the European Union’s Horizon Europe research and innovation programme under the Marie Sk\l{}odowska-Curie grant agreement No 101103830.
THP and JPC acknowledge support from the Agencia Nacional de Investigación y Desarrollo (ANID) grants: CATA-Basal FB210003 and Beca de Doctorado Nacional (JPC).
This research made use of Astropy,\footnote{\url{http://www.astropy.org}} a community-developed core Python package for Astronomy \citep{astropy2013, astropy2018}. This research made use of Photutils, an Astropy package for detection and photometry of astronomical sources \citep{photutils} and Specutils \citep{specutils}, an Astropy package for spectroscopy. Based on observations collected at the European Southern Observatory under ESO programmes 094.B-0612 and 110.23VM.
\end{acknowledgements}

\bibliographystyle{aa} 
\bibliography{References} 

\appendix
\section{Unsuccessful SED fits}
\label{app:bad_SEDs}
\begin{figure}
    \centering
    \includegraphics[width=0.45\textwidth]{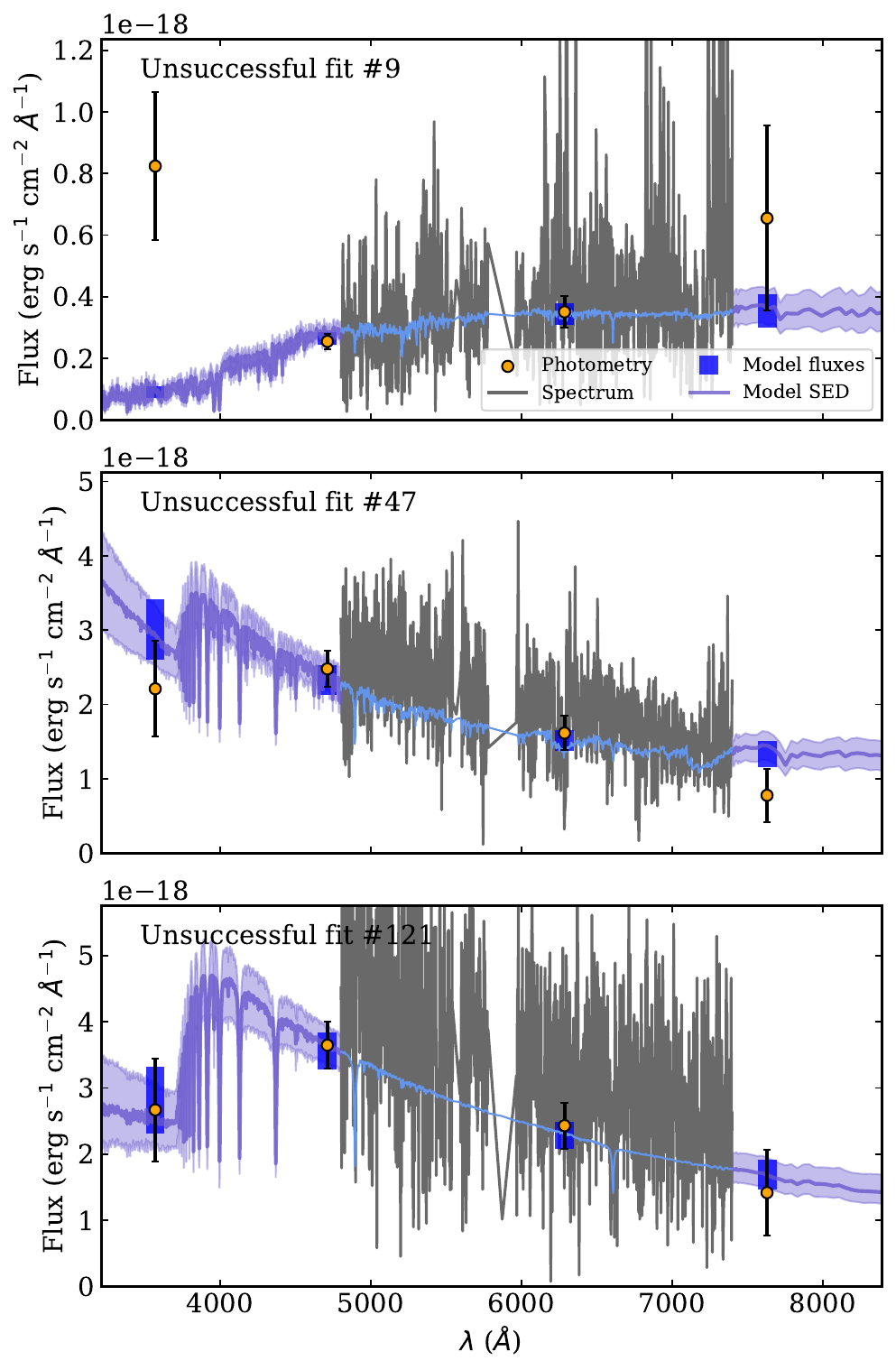}
    \caption{Examples of unsuccessful SED fits. Top panel: Example where neither the spectrum nor the fluxes could be reproduced. Middle and bottom panels: Examples of low $S/N$ spectra, where at least the fluxes could be reproduced.}
    \label{fig:bad_SED_fits}
\end{figure}

Figure \ref{fig:SED_fits} shows three examples of successful SED fits with \textsc{bagpipes}. However, for many sources, the fits were unsuccessful, often due to limited $S/N$ in the spectrum. Examples of those are shown in Fig. \ref{fig:bad_SED_fits}. The top panel shows an example where neither the fluxes nor the spectrum could be fit successfully. This source was removed. The bottom two examples have low $S/N$ spectra that did not enable a spectroscopic fit, however, the fluxes are reasonably well described by the best-fit model. These two sources were included in the photometry-only sample.

\end{document}